\documentstyle[11pt,newpasp,twoside,epsf]{article}
\def\edcomment#1{\iffalse\marginpar{\raggedright\sl#1\/}\else\relax\fi}
\reversemarginpar

\begin{document}
\title{Using Local Group Galaxies to Investigate the Influence
of Blending on Cepheid Distances and the Cosmological Distance Scale}
\author{Barbara J.~Mochejska}
\affil{Harvard-Smithsonian Center for Astrophysics, 60 Garden St,
Cambridge, MA 02138, USA; bmochejs@cfa.harvard.edu}
\author{Lucas M.~Macri}
\affil{National Optical Astronomy Observatories, North Cherry Ave.,
Tucson, AZ 85719, USA; lmacri@noao.edu}
\author{Dimitar D.~Sasselov, Krzysztof Z.~Stanek}
\affil{Harvard-Smithsonian Center for Astrophysics, 60 Garden St,
Cambridge, MA 02138, USA; sasselov@cfa.harvard.edu, kstanek@cfa.harvard.edu}

\begin{abstract}
We investigate the influence of blending on the Cepheid distance scale
using two Local Group galaxies, M31 and M33. Blending leads to
systematically low distances to galaxies observed with the HST, and
therefore to systematically high estimates of $H_0$. High-resolution  
HST images are compared to our ground-based data, obtained as part of
the DIRECT project, for a sample of 22 Cepheids in M31 and 102
Cepheids in M33. For a sample of 22 Cepheids in M31, the average
(median) flux contribution from luminous companions not resolved on
the ground-based images in the $V$-band, $S_V$, is about 19\% (12\%)
of the flux of the Cepheid. For 102 Cepheids in M33 the average
(median) values of $S_V$, $S_I$, $S_B$ are 23\% (13\%), 28\% (20\%),
28\% (15\%). For 64 Cepheids in M33 with periods in excess of 10 days
the average (median) $S_V$, $S_I$, $S_B$ are 16\% (7\%), 23\% (12\%),
20\% (10\%).
\end{abstract}

\begin{figure}[t]
\plotone{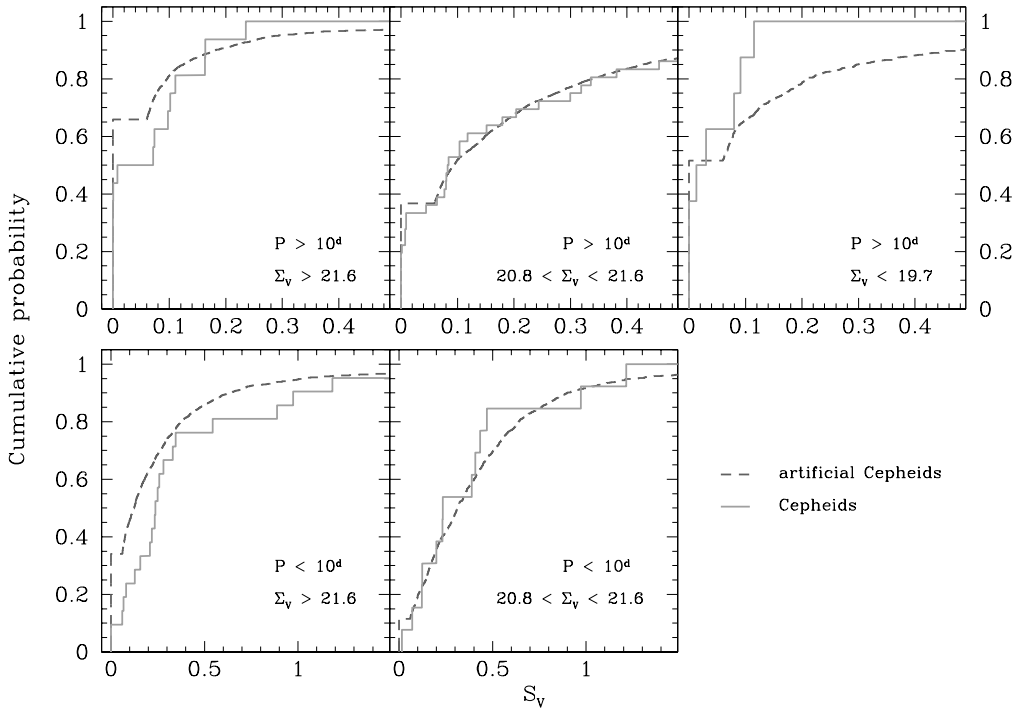}
\caption{The cumulative probability distributions of the blending
parameter $S_V$ for the artificial Cepheids (dashed line) and the
Cepheid catalog (solid).}
\label{fig:ks_test}
\end{figure}

\section{Introduction}
As the number of extragalactic Cepheids discovered with {\em HST}
continues to increase and the value of $H_0$ is sought from distances
based on these variables, it becomes even more important to understand
various possible systematic errors which could affect the
extragalactic distance scale. Currently, the most important systematic
is a bias in the distance to the Large Magellanic Cloud, which
provides the zero-point calibration for the Cepheid distance scale
(e.g. Udalski 2000; Fitzpatrick et al.~2003). Another possible
systematic, the metallicity dependence of the Cepheid
Period-Luminosity (PL) relation, is also very much an open issue, with
empirical determinations ranging from 0 to $-0.4$ mag dex$^{-1}$
(e.g.\ Sasselov et al.~1997; Udalski et al.\ 2001).

We define {\em blending} as the close projected association of a Cepheid
with one or more intrinsically luminous stars, which cannot be detected
within the observed point-spread function (PSF) by photometric analysis.
Blending is thus a phenomenon different from {\em crowding} or {\em
confusion noise}; the latter occurs in stellar fields with a crowded and
complex background due to the random superposition of stars with
different luminosities.

We investigate the effects of stellar blending on the Cepheid distance
scale by studying two Local Group spiral galaxies, M31 and M33.  We
identify some of the Cepheids, discovered by the DIRECT project (Stanek
et al.\ 1999, Mochejska et al.\ 1999) on archival {\em HST}-WFPC2 images
and compare them to our ground-based data to estimate the impact of
blending on our photometry, taking advantage of their superior
resolution.

\section{The Blending Catalogs} We have adopted three criteria that a
companion to a Cepheid has to fulfill to be included into our catalog as
a blend. The star has to: (1) be located at a distance less than
$0\farcs75$ from the Cepheid (half the typical full width at half
maximum on ground-based images), (2) be undetected by DAOPHOT in our
ground-based images, (3) contribute at least 4\% (for M31) or 6\% (for
M33) of the flux of the Cepheid in the same filter.

To quantify blending we have used the parameter $S_F$, defined in
Mochejska et al.\ (2000) as the sum of all flux contributions from
blends in filter $F$ normalized to the flux of the Cepheid: $S_F=
\sum_{i=1}^{N_F}\frac{f_i}{f_C}$ where $f_i$ is the flux of the i-th
blend, $f_C$ the flux of the Cepheid on the {\em HST} image and $N_F$
the total number of blends. Table 1 shows the blending statistics for
M31 and M33. The detailed blending catalogs can be found in Mochejska et
al.\ (2000, 2001). The cumulative probability distribution of blending
in M33 in the V filter is plotted in Figure 1 (continuous line).

\begin{table}
\caption{ Blending Statistics for M31 and M33}
\begin{tabular}{llllllllllllll}
\tableline \tableline
& Period &&     &$S_V$&   &&     &$S_I$&   &&     &$S_B$&   \\
& range  && avg & med & N && avg & med & N && avg & med & N \\
\tableline
M31& all Periods  && 0.19 & 0.12 & 22 &&      &     &    &&      &     &   \\
M33& all Periods  && 0.24 & 0.14 & 95 && 0.30 & 0.21& 62 && 0.29 & 0.15& 57\\
M33& P $<$ 10 days&& 0.37 & 0.25 & 35 && 0.43 & 0.29& 20 && 0.47 & 0.26& 18\\
M33& P $>$ 10 days&& 0.16 & 0.07 & 60 && 0.23 & 0.14& 42 && 0.20 & 0.10& 39\\
\tableline
\end{tabular}
\end{table}

\begin{figure}[t]
\plotfiddle{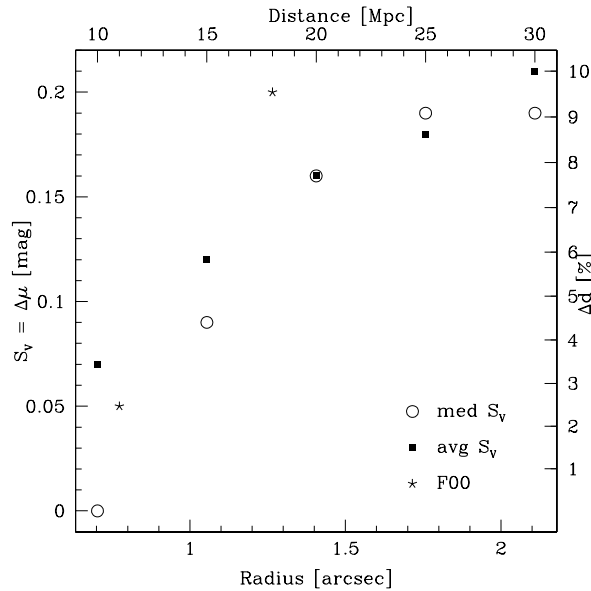}{6.5cm}{0}{40}{40}{-130}{-70}
\caption{Blending/distance bias as a function of the summing
radius/distance. The average and median $S_V$ are indicated with
open and filled symbols, respectively. The Ferrarese et al. (2000)
results are indicated with asterisks.}
\label{fig:bvd}
\end{figure}

\section{ Crowding vs. Blending - Artificial Star Tests}
If we assume that a Cepheid is associated with other luminous stars
located in its proximity, then moving it to a randomly chosen position
on the image will break that association. In the former case, the
Cepheid will be subject to blending; in the latter, to crowding.

To estimate the influence of crowding we have generated an artificial
catalog: for each Cepheid observed on a WFPC2 image we generated a
list of 100 random positions and determined the contribution from
companions at that location.

Figure \ref{fig:ks_test} shows the cumulative probability
distributions for $S_V$ drawn from the artificial crowding catalog
(dashed line) and the Cepheid blending catalog (solid line). We have
divided the sample of M33 Cepheids into two bins in period at $P=10^d$
and three bins in surface brightness, corresponding to regions near
the nucleus, inside the spiral arms and in between them.

For Cepheids located in regions of lowest surface brightness blending
is stronger than crowding. The two effects appear comparable in
magnitude for Cepheids located in intermediate surface brightness
regions. For Cepheids located in the highest surface brightness
regions blending is weaker than crowding, most likely due to selection
effects.

This comparison indicates that the importance of blending relative to
crowding very likely increases with decreasing surface brightness. This
is not unexpected, as young stars are known to cluster (Harris \&
Zaritsky 1999). Increasing the level of crowding will tend to obscure
this effect. 

\section{ Indications for Remote Galaxies}

Using the HST M33 data as the template we have obtained an estimate of
the effect that blending would have on this galaxy if it were observed
at further distances. By increasing the radius around the Cepheid for
summing the contributions from the blends we can simulate the
deterioration of resolution due to the increasing distance to the
galaxy. We have restricted ourselves to the long period Cepheids
($P>10^d$), as they are preferred for determining distances. We have
also rejected all Cepheids with $S_V>45\%$, assuming that they will be
recognized as blended based on the shape of the light curve and rejected
from the sample. The bias in distance due to blending is illustrated in
Fig. \ref{fig:bvd} as a function of distance. The distance underestimate
increases from $0\%-3\%$ to 8\% at 10-15 Mpc and levels off at 9\%-10\%
at 25-30 Mpc. This indicates that blending could potentially be a
substantial source of error in the Cepheid distance scale.

\end{document}